\documentclass[draftcls, onecolumn, 10pt]{IEEEtran}
\usepackage{amssymb,amsmath}
\usepackage[dvips]{graphicx}
\usepackage{amsfonts}
\usepackage{subfigure}
\usepackage{booktabs,multirow}
\usepackage{color}
\usepackage{cite}
\usepackage[table]{xcolor}
\usepackage[justification=centering]{caption}

\IEEEoverridecommandlockouts

\begin{document}

\title{Distinguished Capabilities of Artificial Intelligence Wireless Communication Systems}

\author{\normalsize
Xiaohu Ge,~\IEEEmembership{Senior~Member,~IEEE} \\
\vspace{0.70cm}
\small{
School of Electronic Information and Communications\\
Huazhong University of Science and Technology, Wuhan 430074, Hubei, P. R. China.\\
Email: xhge@mail.hust.edu.cn\\
\vspace{0.2cm}
}
\thanks{\small{ Submitted to IEEE Wireless Communications.}}
\thanks{\small{The authors would like to acknowledge the support from National Key
R\&D Program of China (2016YFE0133000): EU-China study on IoT and 5G (EXICITING-723227). }}
}

\renewcommand{\baselinestretch}{1.2}
\thispagestyle{empty}
\maketitle
\thispagestyle{empty}
\setcounter{page}{1}\begin{abstract}
 With the great success of artificial intelligence (AI) technologies in pattern recognitions and signal processing, it is interesting to introduce AI technologies into wireless communication systems. Currently, most of studies are focused on applying AI technologies for solving old problems, e.g., wireless location accuracy and resource allocation optimization in wireless communication systems. However, It is important to distinguish new capabilities created by AI technologies and rethink wireless communication systems based on AI running schemes. Compared with conventional capabilities of wireless communication systems, three distinguished capabilities, i.e., the cognitive, learning and proactive capabilities are proposed for future AI wireless communication systems. Moreover, an intelligent vehicular communication system is configured to validate the cognitive capability based on AI clustering algorithm. Considering the revolutionary impact of AI technologies on the data, transmission and protocol architecture of wireless communication systems, the future challenges of AI wireless communication systems are analyzed. Driven by new distinguished capabilities of AI wireless communication systems, the new wireless communication theory and functions would indeed emerge in the next round of the wireless communications revolution.

\end{abstract}

\IEEEpeerreviewmaketitle
\newpage
\section{Introduction}
Benefiting from the advance of wireless communication technologies, e.g., the massive multiple-input and multiple-output (MIMO) antennas, millimeter wave transmission and ultra-dense networking technologies, the enhanced mobile broadband (eMBB), ultra-reliable low latency communications (URLLC) and massive machine type communications (mMTC) have been recommended as three typical applications for the fifth generation (5G) wireless communication systems\cite{b1,b2}. However, it is difficult for conventional wireless communication theory and methods to solve all issues of three typical applications, such as massive traffic and accessing, URLLC in 5G wireless communication systems\cite{b3}. With the great development of artificial intelligence (AI) in recent five years, AI technologies are expected to combine into the future wireless communication systems and solve the massive traffic and accessing with ultra-reliable and low latency constraints\cite{b4}. Nevertheless, the revolutionary impact of AI applications on the future wireless communication systems has not been deeply understood. One of important challenges is to indicate the distinguished capabilities and define the features of future AI wireless communication systems.

Currently, the machine learning technology is emerging as one of the most attractive AI technologies for wireless communications. Utilizing the machine learning technology, a robust and efficient algorithm was developed to enhance the accuracy for time-of-arrival localization through identifying and mitigating non-line of sight (NLOS) signals in harsh indoor environments\cite{b5}. By collected and analyzed channel state information (CSI) with a deep learning network with four hidden layers, a deep-learning-based indoor fingerprinting scheme was proposed to improve the location accuracy and reduce the complexity in wireless sensor networks\cite{b6}. Based on the Received Signal Strength Indicator (RSSI) of receivers, three typical learning technologies, e.g., the decision tree, vector machine and neural networks, were compared for improving the accuracy of locations in wireless sensor networks\cite{b7}. Except for the location issue in wireless communications, the AI technologies have been also widely used for the cognitive radio systems\cite{b8,b9}. By categorizing learning problems into the decision-making and feature classification for cognitive radio systems, the working conditions of supervised and unsupervised learning algorithms on the decision-making and feature classification were presented in cognitive radio systems\cite{b8}. To overcome the waveform identification issues in cognitive radio system with high noise environments, the neural network technology was proposed to classify eight typical cognitive radio waveforms\cite{b9}. Driven by the big data in wireless networks, the AI technologies have been investigated to solve the transmission problems of massive traffic for next generation wireless networks\cite{b10,b11,b12}. When the big data need to be transmitted by future wireless networks, AI technologies were used for the data analysis and network efficiency optimization\cite{b10}. To utilize the big data in wireless networks, a data-driven intelligent radio access network architecture was proposed\cite{b11}. Benefits of the next generation wireless networks with machine learning algorithms were presented in \cite{b12}. However, in all the aforementioned studies, AI technologies just have been focused to solve existing issues in wireless communication systems. The new distinguished capabilities created by AI technologies are surprisingly rare for future AI wireless communication systems.

Considering the revolutionary impact of AI technologies on wireless communication systems, three distinguished capabilities of AI wireless communication systems are proposed in this paper. Moreover, an intelligent vehicular communication system is configured as a typical scenario for validating the cognitive capability based on AI clustering algorithm. Based on the results of intelligent vehicular communication systems, the future challenges of AI wireless communication systems are analyzed. All the above factors trigger us to rethink the future AI wireless communication systems in the conclusions.

\section{Capabilities of AI Wireless Communication Systems}
\subsection{AI Learning Algorithms}
AI learning algorithms are ripe for practical applications and great successes have been achieved in the topics of image processing and natural language processing. As a consequence, AI learning algorithms are also expected to apply for wireless communication systems. Based on application objectives, AI learning algorithms are categorized as the supervised and unsupervised, reinforcement learning algorithms.

\begin{itemize}

\item[$\blacktriangleright$] \textbf{Supervised Learning Algorithm}

\end{itemize}

Supervised learning algorithm is a type of AI algorithm in which the mapping function between the input and output can be inferred by the labeled training data. The labelled training data is composed of training pairs which include an input object and a desired output value. The supervised learning algorithm analyzes the training pairs and approximate the mapping function. Based on supervised learning algorithms, the predicted outputs can be inferred by the approximated mapping function and new inputs.

Classification algorithms, e.g., neural network algorithms are a type of supervised learning algorithm in which the program learns from the data input given to it and then uses this learning to classify new observations\cite{b13}. The neural network is composed of a connection of neurons which can process the received signals with non-linear functions. Neurons have a threshold such that the signal is only sent if the aggregate signal crosses the given threshold. The connections between neurons are denoted as edges. Neurons and edges typically have a weight that adjusts as learning proceeds in neural network algorithms. Multiple neurons can form a layer and different layers can transform different outputs by configuring different weights and thresholds on neurons. In generally the signal can be passed through multiple layers with multiple times. Hence, the neural network algorithm is an iterated supervised learning algorithm in which the weights and thresholds are updated in every iteration process. Based on the supervised learning process of neural networks, the massive data of wireless transmissions can be used for data training in neural networks and then obtain the regular pattern of wireless transmissions.

\begin{itemize}

\item[$\blacktriangleright$] \textbf{Unsupervised Learning Algorithm}

\end{itemize}

Unsupervised learning algorithm is a type of AI algorithm in which the underlying structure of data is learned from the data. The data only includes the input without corresponding output values. Therefore, different inputs are utilized to find the relationship or underlying structure in the data. Clustering algorithms are a type of unsupervised learning algorithm which can be used for solving the clustering problem. The core idea of clustering algorithm is to group a set of objects in such a way that objects on the same cluster are more similar to each other than to those in other clusters\cite{b14}. The K-means algorithm is one of the simplest clustering algorithms. K-means algorithm is not only used for the data clustering but also used for adjusting the vehicular network topology. The cluster structure can be used for connecting with vehicles in vehicular communication systems.

\begin{itemize}

\item[$\blacktriangleright$] \textbf{Reinforcement Learning Algorithm}

\end{itemize}

Reinforcement learning algorithm is a type of algorithm in which the agent receives a delayed reward in the next time step to evaluate its previous action. In generally, the reinforcement learning algorithm includes two type of element, i.e., the agent and environment. The agent takes an action based on the initial stage of the environment. Based on the agent action, the environment responds the next stage and reward back to the agent. Furthermore, the agent can update the learning process based on the reward from the environment and then decide the next step action. The iteration process keeps going on until the environment converges to a stationary stage. One of the typical reinforcement learning algorithms is the Q-learning algorithm which is used for solving the fair coexistence issue between long term evolution (LTE) and WiFi communication systems in unlicensed spectrum\cite{b15}.

\subsection{Capabilities of AI Wireless Communication Systems}
In the real world the mobility of terminals, the time-variant wireless channels and the random generated traffic lead the wireless communication system to a complex and dynamic system. The conventional management of wireless communication systems is based on the exact optimal mathematical models, i.e., utilizing the stationary analytical mathematical result to optimize a dynamical wireless communication system. This work model can achieve the local performance optimization but can not conduce to the global optimization in wireless communication systems. When AI learning algorithms are adopted for wireless communication systems, the work model of wireless communication systems will be changed in a revolutionary way. To define the new work model of AI wireless communication systems, the distinguished capabilities created by AI technologies should be first explained. Therefore, in this paper three distinguished capabilities are emerged for AI wireless communication systems which are illustrated in Fig. 1.

\begin{figure*}[!h]
\begin{center}
\includegraphics[width=5in]{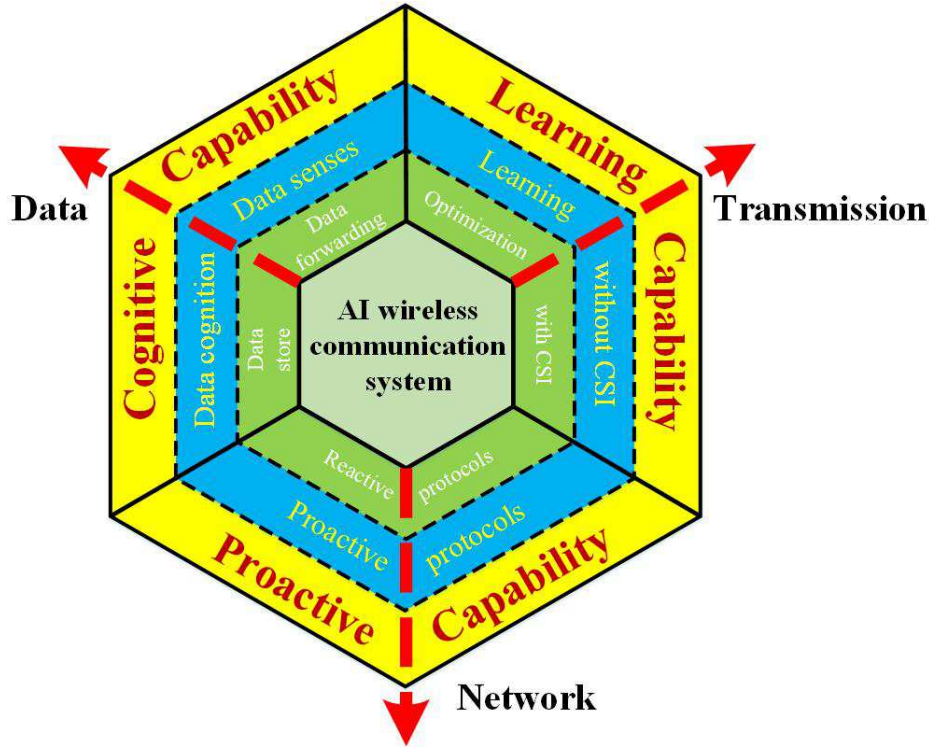}
\caption{Distinguished capabilities of AI wireless communication systems.}\label{Fig1}
\end{center}
\end{figure*}

\begin{enumerate}

\item \textbf{Cognitive Capability:} The data in AI wireless communication systems not only include the application traffic but also include the information of devices and environments in wireless communication systems. The data of conventional wireless communication systems is only used for storing and forwarding. Except for storing and forwarding, the data of AI wireless communication systems is able to be cognized by AI technologies, such as clustering and classification algorithms. The cognitive capability of AI wireless communication systems not only understands the characteristics of data but also senses the generated environment of data. Hence, the cognitive capability is the first distinguishable capability between AI and conventional wireless communication systems.

\item \textbf{Learning Capability:} The conventional wireless transmission process is mainly focused on the wireless link optimization, such as spectral and energy efficiency optimizations. The AI wireless transmission process is recognized as a learning process by AI algorithms. The encoding/decoding and modulating/demodulating are individually designed in conventional wireless communication systems. Moreover, the encoding/decoding and modulating/demodulating is optimized by wireless channel information, e.g., wireless channel state information (CSI). For AI wireless communication systems, the encoding/decoding and modulating/demodulating processes can be processed as different stages of a learning process. For example, the results of encoding and modulating are inputs and the results of decoding and demodulating are outputs for neural networks. The continuous wireless transmission process can be recognized as a continuous learning process in neural networks. In this case, the encoding/decoding and modulating/demodulating can be self-optimized by the learning process without  the wireless channel information if the learning process is enough long for AI wireless transmissions. Therefore, the learning capability is the second distinguished capability between AI and conventional wireless communication systems.

\item \textbf{Proactive Capability:} The protocols of conventional wireless communication systems are a type of reactive protocols. The reactive protocols are composed of fixed functions which are configured in the initial design stage of wireless communication systems. Moreover, the reactive protocols only be triggered by predefined events in wireless communication systems. The protocols of AI wireless communication systems is a type of proactive protocols which can predict and perform in advance to adapt the requirement chances in future wireless communication systems. Therefore, the proactive capability is the third distinguished capability between AI and conventional wireless communication systems.
\end{enumerate}

To validate above distinguished capabilities of AI wireless communication systems, an intelligent vehicular communication system is configured in the Section III. Moreover, the cognitive capability is performed by the AI clustering scheme in the intelligent vehicular communication system.

\section{System Model of Intelligent Vehicular Communications}
\subsection{Framework of intelligent Vehicular Communication Systems}
Conventional vehicular communication systems are designed by an analytical mathematical model which can be optimized by several variables. However, the vehicular communication systems include the high-speed vehicle, the time-variant wireless channels and different application requirements in the real world. All of these factors lead the wireless network topology, wireless transmission capacity and quality of service (QoS) to be coupled each other. Hence, the real vehicular communication system is a complex time-variant system. In most cases, single mathematical model can not describe these complex time-variant relationships of vehicular communication systems. As a consequence, the AI technologies can be applied for vehicular communication systems to fit the complex changes among the wireless network topology, wireless transmission capacity and QoS.

A new intelligent vehicular communication system is illustrated in Fig. 2. Fig. 2(a) shows the intelligent vehicular communication scenario. The intelligent vehicular communication system consists of three types of networks based on different deployment environments. The vehicles on the streets are formed as an ad hoc network which is grouped by different sizes of clusters. The road side units (RSUs) and road side unit centers (RSUCs) along the streets are formed as the wireless access network to connect with vehicles. The cloud service centers (CSCs) and software defined centers (SDCs) are formed as the core network to provide data and resources for vehicles, RSUs and RSUCs.

When vehicles on the streets are grouped to form clusters, vehicles in a cluster are connected by wireless links. One vehicle is dynamically selected as the gateway of cluster to associate with the nearest RSU. Furthermore, all other vehicles in the cluster can connect with the wireless access network by the gateway of cluster. AI technologies can be utilized to form different sizes of intelligent clusters considering the street length, vehicle speed and the signal coverage range of vehicles.

\subsection{Functions of Intelligent Vehicular Communication Systems}
The logical framework of intelligent vehicular communication system is presented at Fig. 2(b), which is composed with three logical layers, i.e., the data cognitive layer, the road side cognitive layer and the vehicle cognitive layer.

\begin{itemize}
\item \textbf{The data cognitive layer} is composed of CSCs and SDCs. The CSC provides the data and resources for requirements of vehicular communications. The SDC manages the resource allocation by AI technologies to proactively match the dynamically changes of wireless channels, vehicular network topology and QoS in intelligent vehicular communication systems.
\item \textbf{The road side cognitive layer} is composed of RSUs and RUSCs. The RSUs take charge of the wireless accessing with vehicles and transmit/receive wireless signals. The RSUCs schedule the resource in adjacent RSUs by AI algorithms to satisfy wireless transmission requirements between RSUs and vehicles.
\item \textbf{The vehicle cognitive layer} is composed of vehicles. Vehicles on the street can be grouped to form clusters. Every cluster has a gateway vehicle which is associated with a RSU. Considering the mobility of vehicles, the gateway vehicle can be dynamically handed over vehicles in the cluster considering the wireless channel capacity between the gateway vehicle and the associated RSU. AI algorithms can be used to dynamically form the cluster considering vehicle numbers and road environments.
\end{itemize}

\begin{figure*}[!h]
\begin{center}
\includegraphics[width=5in]{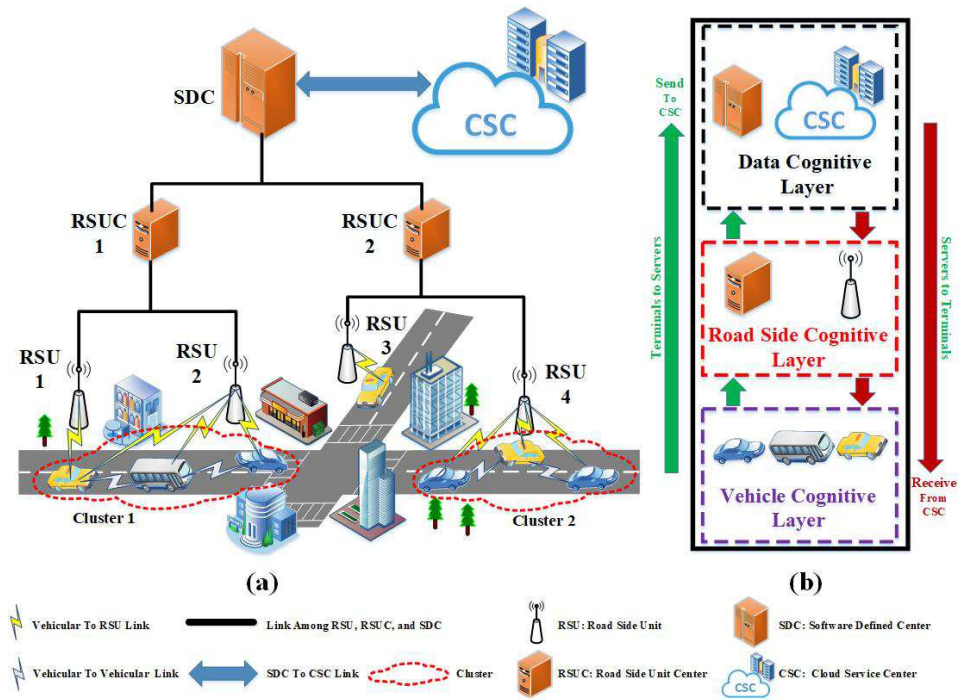}
\caption{Intelligent vehicular communication system. (a) Intelligent vehicular communication scenarios; (b) Logical framework of intelligent vehicular communications.}\label{Fig2}
\end{center}
\end{figure*}

\section{AI Clustering for Intelligent Vehicular Communication Systems}
\subsection{AI Clustering}
Considering the mobility of vehicles, the basic network topologies of vehicular communication systems have to be frequently changed in space and time dimensions. The frequently spatio-temporal variability of vehicular network topology not only increases the overhead of network management but also reduces the data transmission reliability in vehicular communication systems. How to keep the stability of basic network topology is a core issue for vehicular communication systems. Since the mobility of every vehicle can not be controlled by vehicular communication systems, it is impossible to reduce the spatio-temporal variability of vehicle position. However, we can group vehicles into different clusters and keep the stability of cluster structures to reduce the influence of spatio-temporal variability of vehicle positions on the vehicular network topology.

When the millimeter wave transmission technology is adopted for vehicular communication systems, the buildings along streets interrupt the wireless communications among vehicles and RSUs on different streets due to the fast fading of millimeter wave propagations. Considering the sheltering effect of building along streets, the vehicles can not connect with other vehicles on different streets and only can connect with other vehicles in the same straight street. Hence, the number of clusters and the number of vehicles in clusters are depend on the length of straight street. Considering the mobility of vehicles and different coverage ranges of vehicular communication at different vehicles, different clusters are composed of different numbers of vehicles. Moreover, the number of clusters is varied in vehicular communication systems when vehicles turn into different streets.

In conventional vehicular communication systems, the simple environments, e.g., all vehicles are located at the same straight street and have the same coverage range of vehicles, are used for modeling and optimizing the cluster structures. When the vehicles are located at different straight streets and have different wireless coverage ranges of vehicles, it is difficult to optimize the cluster structure of vehicular network topology by a uniform mathematical model. To solve this problem, the AI clustering technology is proposed to form the cluster structure by cognizing environments in intelligent vehicular communication systems.

\subsection{AI Clustering Algorithm of vehicular networks}
In this paper an AI typical clustering algorithm, e.g., the K-means algorithm, is adopted for forming the dynamic cluster structure in intelligent vehicular communication systems. The procedure of K-means algorithm is to distinguish objects into ${k}$ centers. These centers should be placed in a cunning way because different locations cause different results. The better choice is to place them as much as possible far away from each other. The next step is to take each object belonging to a given data set and associate it to the nearest center. When no object is pending, the first step is completed and an early group age is done. At this stage ${k}$ new centroids need to be calculated as the barycenter of the centers resulting from the previous steps. After ${k}$ new centroids are obtained, a new binding has to be associated between the same data set and the nearest new center. The above process is iterated until ${k}$ centers locations do not change any more.

Based on the locations of vehicles on the road, the vehicle cognitive layer is assumed to obtain the density of vehicles, length of straight streets and wireless communication coverage radius of every vehicle. The obtained vehicles and streets information are input for the K-means algorithm in the vehicle cognitive layer. The K-means algorithm continues to generate different number of clusters with different vehicles. The vehicle cognitive layer estimates whether the generated cluster structures, i.e., the number of clusters and the number of vehicles in every cluster and the gateways of every cluster, satisfy wireless linking conditions in intelligent vehicular communication systems. When the generated cluster structures satisfy the wireless linking conditions, the vehicle cognitive layer outputs the result of cluster structure to all vehicles on streets. The AI clustering algorithm flowchart is illustrated in Fig. 3.

Based on the proposed AI clustering algorithm, we can evaluate the connection probability of intelligent vehicular communication systems. When the transmission distance threshold of vehicle is configured as ${R}$ and the vehicle density on the road is ${\rho}$, the connection probability of vehicle ${{{V}_{i}}}$ is ${P=P({{r}_{i}}\le R)=1-{{e}^{-\rho R}}}$, where ${{{r}_{i}}}$ is the wireless communication coverage radius of vehicle ${{{V}_{i}}}$. Based on the AI cluster structure, the gateway vehicle can cover all vehicles in the AI cluster. Assume that the gateway vehicle and other vehicles in the cluster have the same transmission distance threshold ${R}$. The transmission distance threshold of AI cluster is configured as ${2R}$. There are ${n}$ vehicles on the straight street and ${k}$ clusters have been grouped by the K-means algorithm. Assumed that ${m}$ vehicles have not been grouped into any clusters in the road. The cluster probability is ${q={}^{k}/{}_{n}}$ and the single vehicle probability is ${1-q}$. Hence, the connection probability of vehicular communication system on the straight street is ${{{P}_{c}}={{[(1-q)(1-{{e}^{-\rho R}})+q(1-{{e}^{-2\rho R}})]}^{m+k}}}$.

\begin{figure*}[!h]
\begin{center}
\includegraphics[width=5in]{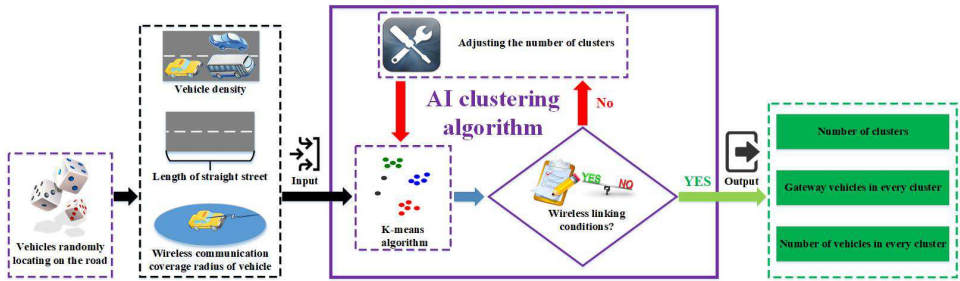}
\caption{AI clustering algorithm flowchart.}\label{Fig3}
\end{center}
\end{figure*}

\subsection{Performance Analysis of AI Clustering Algorithm}
In this section simulations are performed to compare connect probabilities of vehicular communication systems with and without AI clustering algorithm. Without loss of generality, locations of vehicles are assumed to be governed by a Poisson distribution with the vehicle density ${\rho =0.1}$ vehicle per meter.

When the AI clustering algorithm is adopted for intelligent vehicular communication systems, Fig. 4 shows the optimized number of clusters with respect to the length of straight street and the wireless communication coverage radius of vehicle. When the length of straight street is fixed, the optimized number of clusters decreases with the increase of the wireless communication coverage radius of vehicle. When the wireless communication coverage radius of vehicle is fixed, the optimized number of clusters increases with the increase of the length of straight street.

Considering urban environments, 5 kilometers road is configured to analyze the performance of vehicular communication systems with and without AI clustering algorithm. In Fig. 5, 5 kilometers road is composed with five straight streets with different lengths, i.e., the 600, 800, 1000, 1200, 1400 meters straight streets. Every cluster only can be located on one of five straight streets. To compare with the AI clustering algorithm, a non-cluster algorithm, i.e., every vehicle directly connects with RSUs along the street is plotted in Fig. 5. Based on the results in Fig. 5, the connection probability of intelligent vehicular communication systems with AI clustering algorithm is larger than the connection probability of vehicular communication systems with the non-cluster algorithm. This result implies that the AI clustering algorithm can improve the stability of vehicular network topology by increasing the connection probability of intelligent vehicular communication systems.

\begin{figure*}[!h]
\begin{center}
\includegraphics[width=5in]{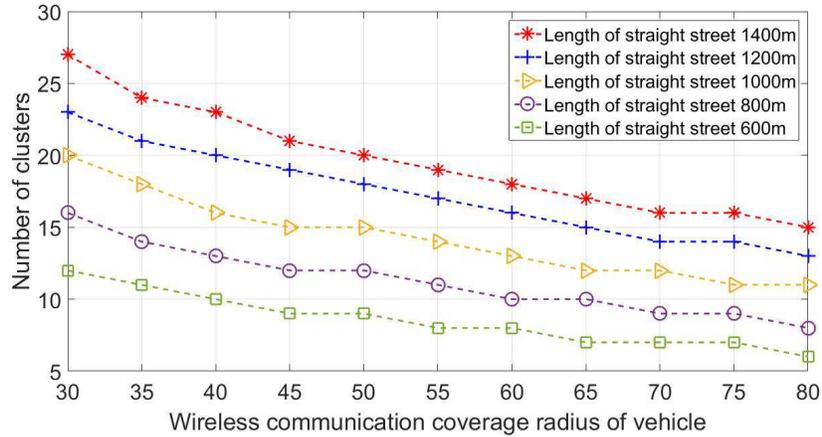}
\caption{Optimized number of clusters with respect to the length of straight street and the wireless communication coverage radius of vehicle.}\label{Fig4}
\end{center}
\end{figure*}

\begin{figure*}[!h]
\begin{center}
\includegraphics[width=5in]{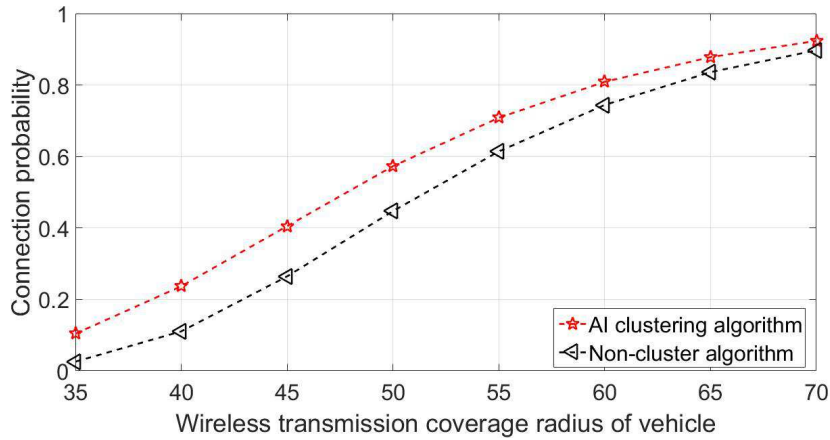}
\caption{Connection probability of vehicular communication systems with and without AI clustering algorithm.}\label{Fig5}
\end{center}
\end{figure*}

\section{Challenges of Intelligent Wireless Communication Systems}
When the cognitive capability, e.g., the AI clustering scheme is adopted, the impact of vehicle mobility on the stability of vehicular network topology can be reduced and the connect probability of intelligent vehicular communication systems is improved. The cognitive, learning and proactive capabilities not only improve the performance but also change the future work models of AI wireless communication systems. Hence, some potential challenges of AI wireless communication systems are summarized as follows:

The first challenge is AI technologies for data in AI wireless communication systems. In conventional wireless communication systems, the data is generated from application traffic and individually stored and computed in different devices. In AI wireless communication systems, the data not only includes the application traffic but also comes from the information of devices and environments. Considering the data generated from different types of sources, it is an important issue how to express different types of data by a flexible and scalable form which can be used for different AI algorithms in AI wireless communication systems. When AI supervised learning algorithms are adopted for the cognitive capability, it is a great challenge to label the massive training data in AI wireless communication systems.

The second challenge is the transmission in AI vehicular communication systems. In conventional wireless communication systems, the transmission process is a coding and modulation process. Moreover, the wireless transmission is focused on the performance optimization based on the CSI of wireless channels and resource distribution information. For AI vehicular communication systems, the transmission process is regarded as a learning process. Based on the massive transmission data and the continuous data transmission process, the transmitters and receivers can be configured as the input and output of AI iterated algorithms to optimize the transmission efficiency without CSI and resource distribution information. It is a new challenge to redesign the coding and modulation schemes based on AI algorithm in AI vehicular communication systems.

The third challenge is the AI technologies for the protocol architecture of AI wireless communication systems. The conventional protocol architecture of wireless communication systems is a layered protocol architecture, which is help for the exact function structure in distributed devices. However, it is difficult for the layered protocol architecture to perform an un-exact function structure in AI wireless communication systems. For example, when the cognitive and learning capabilities are performed by neural network or deep learning technologies in AI wireless communication systems, we do not know how many layers should be configured for neural network or deep learning technologies. In this case, it is difficult to deploy the neural network or deep learning technologies in distributed devices based the conventional layered protocol architecture. To achieve the cognitive, learning and proactive capabilities for AI wireless communication systems, a new intelligent protocol architecture must be investigated based on AI running schemes. The new intelligent protocol architecture should support the AI technologies to running in distributed devices. It is a great challenge to combine the AI running schemes into the layered protocol architecture and support cognitive, learning and proactive capabilities in AI wireless communication systems.

\section{Conclusion}
In this paper we analyze prospects of AI technologies for wireless communication systems. Based on the revolutionary impact of AI technologies, three distinguished capabilities, i.e., the cognitive, learning and proactive capabilities are proposed for future AI wireless communication systems. To validate the cognitive capability based on AI technologies, an intelligent vehicular communication system is configured as a typical scenario for AI applications. Moreover, an AI clustering algorithm based on K-means algorithm is proposed for intelligent vehicular communication systems. Simulation results indicate that the proposed AI clustering algorithm can improve the connection probability of intelligent vehicular communication systems. Based on the results of AI clustering algorithm, the challenges of AI wireless communication systems, i.e., the AI technologies for data, transmission and protocol architecture of AI wireless communication systems are analyzed. When above challenges are realized for future AI wireless communication systems, the revolutionary chances will bring us into a new information world.

\end{document}